\documentstyle[aps,psfig]{revtex}
\begin{document}
\draft
\title{Effect of $q$-deformation in the NJL gap equation}
\author{V. S. Tim\'{o}teo and C. L. Lima}
\address{Nuclear Theory and Elementary Particle Phenomenology Group\\
{\em Instituto de F\'{\i }sica, Universidade de S\~{a}o Paulo}\\
{\em CP 66318, 05315-970, S\~{a}o Paulo, SP, Brazil}}
\maketitle

\begin{abstract}
We obtain a $q$-deformed algebra version of the Nambu--Jona-Lasinio model
gap equation. In this framework we discuss some hadronic properties such as
the dynamical mass generated for the quarks, the pion decay constant and the
phase transition present in this model.
\end{abstract}

\date{\today }

The concept of symmetry is of fundamental importance in physics; the
breaking of a symmetry and its associated phase transition are universal
phenomena appearing in many branches in physics, such as nuclear and solid
state physics, although the broken symmetries and the physical systems
involved are quite different. Y. Nambu was the first to realize this
universal aspect of dynamical symmetry breaking \cite{NJL}. The
Nambu-Jona-Lasinio (NJL) model is very adequate to study the breaking of
chiral symmetry and the generation of a dynamical mass for the quarks due to
the appearance of condensates.

On another side, in the last few years the study of $q$-deformed algebras
turned out to be a fertile area of research. The use of $q$-deformed algebra
in the description of some many-body systems has lead to the appearance of
new features when compared to the non-deformed case \cite{GLPL}. In
particular, it seems to be a very elegant framework to describe
perturbations from some underlying symmetry structure. From the many
applications of $q$-deformation ideas existing in the literature, ranging
from optics to particle physics, we would like to pinpoint three of them:
the investigation of the behavior of the second order phase transition in a $%
q$-deformed Lipkin model \cite{gal}, the good agreement with the
experimental data obtained through a $\kappa $-deformed Poincar\'{e}
phenomenological fit to the dynamical mass and rotational and radial
excitations of mesons \cite{day}, and the purely $su_q(2)$-based mass
formula for quarks and leptons developed by using an inequivalent
representation \cite{pala}.

It sounds therefore reasonable to study the influence of the $q$-deformation
on the mass generation mechanism due to the breaking of chiral symmetry. To
be definite, in this work we intend to investigate the effects of the $q$%
-deformation on the phase transition of the NJL model, stimulated by an
analogy with $q$-deformed Lipkin model, where the phase transition is
smoothed down when the Lipkin Hamiltonian is deformed \cite{gal}.

Recently, the thermodynamical properties of a free quantum group fermionic
system with two ``flavors'' were studied \cite{ubri}. In particular, it was
given there a $su_q\left( N\right) $-covariant representation of the
fermionic algebra for arbitrary $N$ in terms of ordinary creation and
annihilation operators. The $su_q\left( 2\right) $-covariant algebra is
given by the following relations 
\begin{equation}
\left\{ \psi _1,\overline{\psi }_1\right\} =1-\left( 1-q^{-2}\right) 
\overline{\psi }_2\psi _2\qquad \left\{ \psi _2,\overline{\psi }_2\right\}
=1,
\end{equation}
\begin{equation}
\psi _1\psi _2=-q\psi _2\psi _1\qquad \overline{\psi }_1\psi _2=-q\psi _2%
\overline{\psi }_1,
\end{equation}
\begin{equation}
\left\{ \psi _1,\psi _1\right\} =0\qquad \left\{ \psi _2,\psi _2\right\} =0.
\end{equation}
The usual $su(2)$ covariant fermionic algebra is recovered when $q=1$.
Later, the pure nuclear pairing force version of the
Bardeen-Cooper-Schrieffer (BCS) many-body formalism \cite{ring-schuck} was
extended in such a way to replace the usual fermions by quantum group
covariant ones satisfying appropriate anticommutation relations for a $%
su_q(N)$-fermionic algebra \cite{trip}. Using the $su_q\left( 2j+1\right) $%
-covariant representation of the fermionic algebra, a $q$-covariant form of
the BCS approximation was constructed and the $q$-analog to the BCS
equations along with the quantum gap equation was derived. The quantum gap
was shown to depend explicitly on the deformation parameter and it is
reduced as the deformation increases.

The Nambu--Jona-Lasinio model was first introduced to describe the
nucleon-nucleon interaction via a four-fermion contact interaction. Later,
the model was extended to quark degrees of freedom becoming an effective
model for quantum chromodynamics (QCD).

The Lagrangian of the NJL model is given by 
\begin{eqnarray}
{\cal L}_{NJL} &=&\overline{\psi }i\gamma ^\mu \partial _\mu \psi +{\cal L}%
_{int},  \label{lag} \\
{\cal L}_{int} &=&G\left[ \left( \overline{\psi }\psi \right) ^2+\left( 
\overline{\psi }i\gamma _5{\bf \tau }\psi \right) ^2\right] .  \label{int}
\end{eqnarray}
Linearizing the above interaction in a mean field approach, the last term
does not contribute if the vacuum is parity and Lorentz invariant. The
Lagrangian with the linearized interaction is then 
\begin{equation}
{\cal L}_{NJL}=\overline{\psi }i\gamma ^\mu \partial _\mu \psi
+2G\left\langle \overline{\psi }\psi \right\rangle \overline{\psi }\psi .
\label{lin}
\end{equation}
Regarding this Lagrangian as a Dirac Lagrangian for massive quarks we obtain
a dynamical mass for the quarks 
\begin{equation}
m=-2G\left\langle \overline{\psi }\psi \right\rangle ,  \label{mdy}
\end{equation}
where $\left\langle \overline{\psi }\psi \right\rangle $ is the vacuum
expectation value of the scalar density $\overline{\psi }\psi $,
representing the quark condensates. Eq. (\ref{mdy}) describes how the
dynamical mass is generated with the appearance of the quark condensates.
The quarks are massless if the condensate vanishes.

We now turn to the $q$-deformation of the NJL gap equation. Following \cite
{ubri,trip} we write the creation and annihilation operators of the $%
su_q\left( 2j+1\right) $ fermionic algebra as, 
\begin{eqnarray}
A_{jm} &=&a_{jm}\prod_{i=m+1}\left( 1+Qa_{ji}^{\dagger }a_{ji}\right) , \\
A_{jm}^{\dagger } &=&a_{jm}^{\dagger }\prod_{i=m+1}\left( 1+Qa_{ji}^{\dagger
}a_{ji}\right) ,
\end{eqnarray}
where $Q=q^{-1}-1,$ $j=1/2$ and $m=\pm 1/2$. The first consequence of the
above deformation is that only the operators corresponding to $m=-\frac 12$
are modified. Since in the NJL model we deal with quarks (anti-quarks)
creation and annihilation operators, this feature is important because only
negative helicity quarks (anti-quarks) operators will be deformed.
Explicitly, we have
\begin{equation}
A_{-}=a_{-}\left(1+Qa_{+}^{\dagger }a_{+}\right) \;\;\; , \;\;\;
A_{-}^{\dagger}=a_{-}^{\dagger  }\left(1+Qa_{+}^{\dagger }a_{+}\right),
\label{am}
\end{equation} 
\begin{equation}
A_{+}=a_{+} \;\;\; , \;\;\; A_{+}^{\dagger }=a_{+}^{\dagger },  
\label{ap}
\end{equation}
where $+$ $(-)$ stands for the positive (negative) helicity. In a sense, we
are embedding the chiral symmetry breaking effects in the operators'
definition.

We are now in position a to obtain the deformed gap equation by introducing a
BCS-like vacuum and proceeding similarly to the standard Bogoliubov-Valatin
approach \cite{K}. The $q$-deformed BCS\ vacuum reads 
\begin{equation}
\left| NJL\right\rangle =\prod_{{\bf p},s=\pm 1}\left[ \cos \theta (p)+s\sin
\theta (p)B^{\dagger }({\bf p},s)D^{\dagger }(-{\bf p},s)\right] \left|
0\right\rangle  \label{bcs}
\end{equation}
and the quark fields are expressed in terms of $q${\em -}deformed creation
and annihilation operators as 
\begin{equation}
\psi _q(x,0)=\sum_s\int \frac{d^3p}{\left( 2\pi \right) ^3}
\left[ B({\bf p},s)u({\bf p},s)e^{i{\bf p\cdot x}}+D^{\dagger }({\bf p},s)
v({\bf p},s)e^{-i{\bf p\cdot x}}\right] .  \label{bd}
\end{equation}
The $q$-deformed quark and anti-quark creation and annihilation operators $B$,
$B^{\dagger }$, $D$, and $D^{\dagger }$, are expressed in terms of
the non-deformed ones according to Eqs. (\ref{am}) and (\ref{ap})
\begin{eqnarray}
B_{-} &=& b_{-}\left(1+Qb_{+}^{\dagger }b_{+}\right) \;\;\; , \;\;\;
B_{-}^{\dagger}=b_{-}^{\dagger  }\left(1+Qb_{+}^{\dagger }b_{+}\right), \\
D_{-} &=& d_{-}\left(1+Qd_{+}^{\dagger }d_{+}\right) \;\;\; , \;\;\;
D_{-}^{\dagger}=d_{-}^{\dagger  }\left(1+Qd_{+}^{\dagger }d_{+}\right),
\label{bm}
\end{eqnarray}  
\begin{eqnarray} 
B_{+} &=& b_{+} \;\;\; , \;\;\; B_{+}^{\dagger }=b_{+}^{\dagger }, \\
D_{+} &=& d_{+} \;\;\; , \;\;\; D_{+}^{\dagger }=d_{+}^{\dagger }, 
\label{bp} 
\end{eqnarray}  
( in the above equations we simplified the notation: 
$B({\bf p},s)\rightarrow B_s$, $b({\bf p},s)\rightarrow b_s$, etc. ).
We would like to stress that, as discussed
 in Ref. \cite{trip},
the deformed vacuum differs from the non-deformed one
 only by a phase and,
therefore, the effects of the deformation comes solely
 from the modified
field operators. Additionally, the $q$-deformed NJL Lagrangian, constructed
using $\psi_q$ instead of $\psi$, is invariant under the quantum group
$SU_q(2)$ transformations. This can be seen by using the two-dimensional 
representation of the $SU_q(2)$ unitary transformation given in 
Ref. \cite{ubri}.

The deformed gap equation is 
\begin{equation}
m=-2G\left\langle \overline{\psi }\psi \right\rangle _q,  \label{qmdy}
\end{equation}
where $\left\langle \overline{\psi }\psi \right\rangle _q$ is the $q$%
-deformed condensate calculated using the BCS-like vacuum, Eq. (\ref{bcs}),
and Eq. (\ref{bd}), 
\begin{eqnarray}
\left\langle \overline{\psi }\psi \right\rangle _q &=&\left\langle NJL\left| 
\bar{\psi}_q\psi _q\right| NJL\right\rangle  \nonumber \\
&=&\left\langle \overline{\psi }\psi \right\rangle +\left\langle NJL\left| 
{\cal Q}\right| NJL\right\rangle ,
\end{eqnarray}
where $\left\langle \overline{\psi }\psi \right\rangle $ is the non-deformed
condensate and $\left\langle NJL\left| {\cal Q}\right| NJL\right\rangle $
represents all non-vanishing matrix elements arising from the $q$%
-deformation of the quark fields. The contribution of these $q$-deformed
matrix elements is 
\begin{equation}
\left\langle NJL\left| {\cal Q}\right| NJL\right\rangle =Q\int \frac{d^3p}{%
\left( 2\pi \right) ^3}\left[ \sin 2\theta (p)-\sin 2\theta (p)\cos 2\theta
(p)\right] .
\end{equation}

The calculation of the deformed condensate will be performed in a similar
way as in the usual case \cite{VW}. It requires also a regularization
procedure since the NJL interaction is not perturbatively renormalizable.
For reasons of simplicity a non-covariant trimomentum cutoff is applied
arising 
\begin{equation}
\left\langle \overline{\psi }\psi \right\rangle _q=-\frac{3m}{\pi ^2}\left[
\left( 1-\frac Q2\right) \int_0^\Lambda dp\frac{p^2}{\sqrt{{\bf p}
^2+m^2}}+\frac Q2\int_0^\Lambda dp\frac{p^3}{{\bf p}^2+m^2}\right]
\label{qq}
\end{equation}
for each quark flavor. At this point we see that the dynamical mass is again
given by a self-consistent equation since the condensate depends also on the
mass. Inserting Eq. (\ref{qq}) into Eq. (\ref{qmdy}) we obtain the deformed
NJL\ gap equation 
\begin{equation}
m=\frac{2Gm}{\pi ^2}\left[ \left( 1-\frac Q2\right) \int_0^\Lambda dp
\frac{p^2}{\sqrt{{\bf p}^2+m^2}}+\frac Q2\int_0^\Lambda dp
\frac{p^3}{{\bf p}^2+m^2}\right] .  \label{qgap}
\end{equation}
It is easy to see that for $Q=0$ $(q=1)$, we recover the NJL gap equation in
its more familiar form 
\begin{equation}
m=\frac{2Gm}{\pi ^2}\int_0^\Lambda dp
\frac{p^2}{\sqrt{{\bf p}^2+m^2}} +m_0,  \label{gap}
\end{equation}
where $m_0$ appears only if we consider the current quark mass term ${\cal L}%
_{mass}=-m_0\overline{\psi }\psi $ in the NJL Lagrangian Eq. (\ref{lag}).

The pion decay constant is calculated from the vacuum to one pion axial
vector current matrix element, which, in the simple 3D non-covariant cutoff
we are using \cite{K}, is given by 
\begin{equation}
f_\pi ^2=N_c m^2\int_0^\Lambda \frac{d^3p}{\left( 2\pi \right) ^3}
\frac 1{\left( {\bf p}^2+m^2\right) ^{3/2}},  \label{qfpi}
\end{equation}
for each quark color. The deformed calculation of $f_\pi $ is performed
directly by substituting the dynamical mass in Eq. (\ref{qfpi}) from the one
obtained in Eq. (\ref{qgap}), instead of deforming the axial current in the
calculation of its matrix element of between the vacuum and the one pion
state.

\begin{figure}[tbh]
\centerline{\psfig{figure=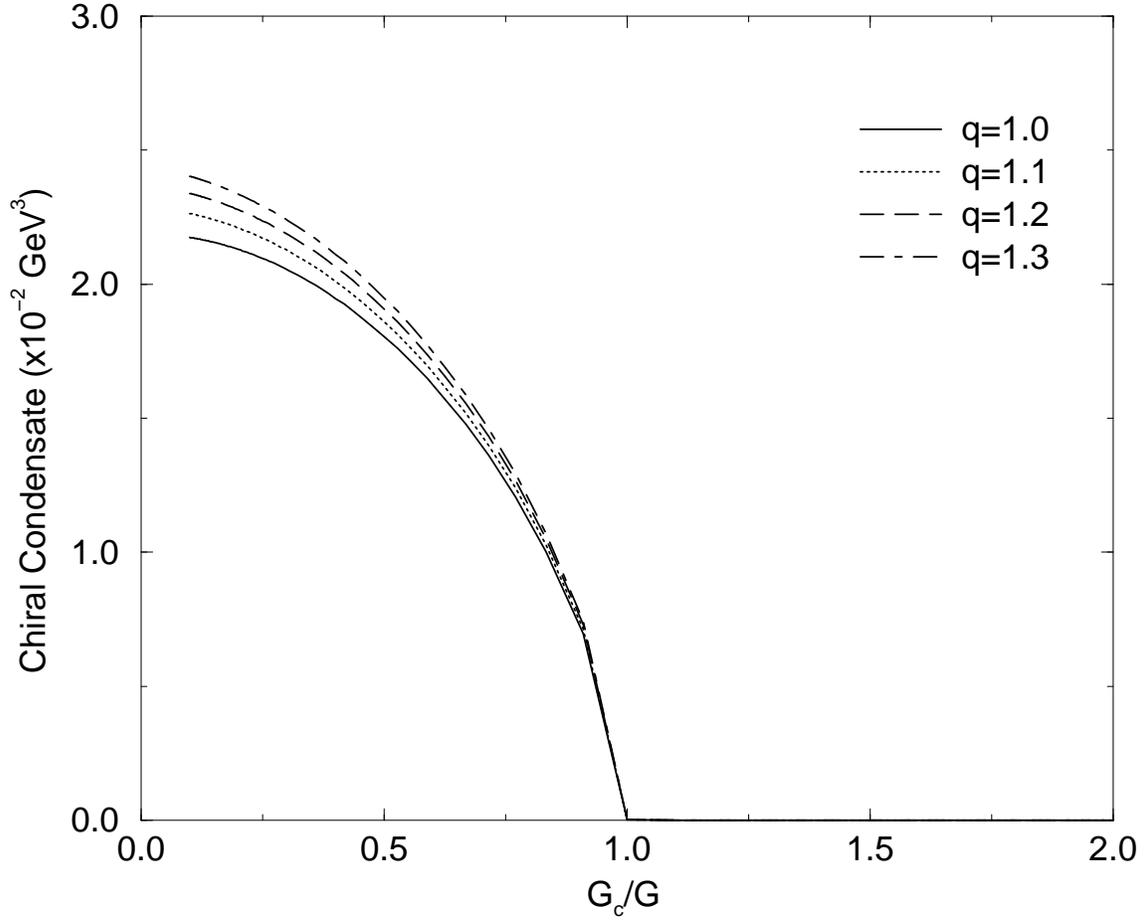,width=15cm}}
\caption{Behavior of the phase transition for different values of $q.$ The
non-deformed situation corresponds to $q=1$. In all curves $\Lambda =600MeV$
and $m_0=0$.}
\label{qpt}
\end{figure}

As in the non-deformed case, the $q$-gap equation has non-trivial solutions
when the coupling $G$ exceeds a critical value $G_{crit}$ related to the
cutoff. Figure \ref{qpt} depicts the sharp phase transition at $G=G_{crit}$\
separating the Wigner-Weyl and Nambu-Goldstone phases, corresponding to
different realizations of chiral symmetry.

Figure \ref{qpt} also shows the deformed condensate values as a function of $%
q$. We can see the enhancement of the condensate's value, due to presence of
the $q$-deformation. The dynamical mass is accordingly modified through the
deformed gap equation (\ref{qgap}), as can be seen in Table \ref{tab1},
along with the corresponding values of the pion decay constant, $f_\pi $.
The behavior of the condensate around the critical coupling, $G_c,$ is
similar for both deformed and non-deformed cases$,$ meaning that the adopted
procedure to $q$-deform the underlying $su(2)$ algebra in a two flavor NJL
model does not change the behavior of the phase transition around $G_c$.
Table \ref{tab2} presents the behavior of the coupling constant for two
typical dynamical mass values for different $q$'s$.$ The analysis of this
table shows that the coupling constant $G$\ decreases with $q$, for a given
value of the dynamical mass, meaning that to acquire a given mass we need a
weaker coupling when the algebra is deformed. This indicates that the
deformation of the $su(2)$\ algebra incorporates effects the NJL interaction
which are propagated to the physical quantities such as the condensate, the
dynamical mass and the pion decay constant.
The formalism developed in
\cite{ubri,trip} allow $q$-values smaller than one (which corresponds to 
$Q > 0$). It is worth to mention that in this case the $q$-deformation effect 
goes in the opposite direction, namely, the condensate value and the dynamical
mass decrease for $q < 1$.
 
To summarize, the main objective of this work was to analyze the influence
of the $q$-deformation in the NJL model. We studied the deformation of the
underlying $su(2)$\ algebra in a two flavor version of the model and
investigated an important feature of the $su(2)$ chiral symmetry breaking,
namely the dynamical mass generation, through the incorporation of helicity
non-conserving terms directly in the fermionic operators. The main effect of
the $q$-deformation is to effectively enhance the coupling strength of the
NJL four fermion interaction, leading to an increasing in the value of the
quark condensate. The dynamical mass, which is related to the presence of
the condensate, is correspondingly increased. We also looked closely at the
behavior of the phase transition around the critical point, which is still
sharp, meaning that the new contributions arising from the deformation of the
condensate do not play the role of explicit chiral symmetry breaking terms \cite
{VW}.

{\bf Acknowledgments }

The authors are grateful to D. Galetti and B. M. Pimentel for helpful
discussions. V. S. T. would like to acknowledge Funda\c {c}\~{a}o de Amparo
\`{a} Pesquisa do Estado de S\~{a}o Paulo (FAPESP) for financial support and
M. Malheiro for very helpful discussions concerning the $q$-deformation in
the NJL model.

\begin{table}[tbp] \centering
\begin{tabular}{ccccc}
& ${\bf q=1.0}${\bf \ } & ${\bf q=1.1}$ & ${\bf q=1.2}$ & ${\bf q=1.3}$ \\ 
&  &  &  &  \\ 
${\bf Mass}$ ${\bf [MeV]}$ & 365 & 375 & 384 & 392 \\ 
&  &  &  &  \\ 
${\bf f}_{\pi }${\bf \ }${\bf [MeV]}$ & 92.0 & 92.4 & 92.9 & 93.3 \\ 
&  &  &  & 
\end{tabular}
\caption{Mass and $f_\pi$ for different values of $q$, for $\Lambda$=600 MeV, 
$G_c$=4.57 GeV$^{-2}$, and $G$=6.53 GeV$^{-2}$. The condensate was calculated
for two flavors and three colors. \label{tab1}}%
\end{table}

\begin{table}[tbp] \centering
\begin{tabular}{ccccc}
${\bf Mass}$ ${\bf [MeV]}$ & ${\bf q=1.0}${\bf \ } & ${\bf q=1.1}$ & ${\bf %
q=1.2}$ & ${\bf q=1.3}$ \\ 
&  &  &  &  \\ 
${\bf 300}$ & 6.04 & 5.98 & 5.94 & 5.90 \\ 
&  &  &  &  \\ 
${\bf 350}$ & 6.42 & 6.35 & 6.28 & 6.23 \\ 
&  &  &  & 
\end{tabular}
\caption{Behavior of the coupling constant $G$ at given values of the dynamical 
mass for different values of $q$. Cutoff and critical coupling are the same as in 
Table \ref{tab1}. \label{tab2}}%
\end{table}

\end{document}